# Astro2020 Science White Paper

# Insights Into the Epoch of Reionization with the Highly-Redshifted 21-cm Line

**Thematic Areas:**  ☐ Planetary Systems   ☐ Star and Planet Formation
☐ Formation and Evolution of Compact Objects   ☒ **Cosmology and Fundamental Physics**
☐ Stars and Stellar Evolution   ☐ Resolved Stellar Populations and their Environments
☒ Galaxy Evolution   ☐ Multi-Messenger Astronomy and Astrophysics

**Principal Author:**
Name: Steven Furlanetto
Institution: University of California, Los Angeles
Email: sfurlane@astro.ucla.edu
Phone: (310) 206-4127

**Co-authors:**
Chris L. Carilli (Cavendish Laboratory, NRAO), Jordan Mirocha (McGill University), James Aguirre (University of Pennsylvania), Yacine Ali-Haimoud (New York University), Marcelo Alvarez(University of California Berkeley), Adam Beardsley (Arizona State University), George Becker (University of California Riverside), Judd D. Bowman (Arizona State University), Patrick Breysse (CITA), Volker Bromm (University of Texas at Austin), Philip Bull (Queen Mary University of London), Jack Burns (University of Colorado Boulder), Isabella P. Carucci (University College London), Tzu-Ching Chang (JPL), Xuelei Chen (National Astronomical Observatory of China), Hsin Chiang (McGill University), Joanne Cohn (University of California Berkeley), Frederick Davies (University of California Santa Barbara), David DeBoer (University of California Berkeley), Joshua Dillon (University of California Berkeley), Olivier Doré (JPL, California Institute of Technology), Cora Dvorkin (Harvard University), Anastasia Fialkov (University of Sussex), Nick Gnedin (Fermilab), Bryna Hazelton (University of Washington), Daniel Jacobs (Arizona State University), Kirit Karkare (University of Chicago/KICP), Saul Kohn (The Vanguard Group), Leon Koopmans (Kapteyn Astronomical Institute), Ely Kovetz (Ben-Gurion University), Paul La Plante (University of Pennsylvania), Adam Lidz (University of Pennsylvania), Adrian Liu (McGill University), Yin-Zhe Ma (University of KwaZulu-Natal), Yi Mao (Tsinghua University), Kiyoshi Masui (MIT Kavli Institute for Astrophysics and Space Research), Matthew McQuinn (University of Washington), Andrei Mesinger (Scuola Normale Superiore),

Julian Munoz (Harvard University), Steven Murray (Arizona State University), Aaron Parsons (University of California Berkeley), Jonathan Pober (Brown University), Benjamin Saliwanchik (Yale University), Jonathan Sievers (McGill University), Eric Switzer (NASA Goddard Space Flight Center), Nithyanandan Thyagarajan (NRAO), Hy Trac (Carnegie Mellon University), Eli Visbal (Flatiron Institute), Matias Zaldarriaga (Institute for Advanced Study)

**Abstract**: The epoch of reionization, when photons from early galaxies ionized the intergalactic medium about a billion years after the Big Bang, is the last major phase transition in the Universe's history. Measuring the characteristics of the transition is important for understanding early galaxies and the cosmic web and for modeling dwarf galaxies in the later Universe. But such measurements require probes of the intergalactic medium itself. Here we describe how the 21-cm line of neutral hydrogen provides a powerful probe of the reionization process and therefore important constraints on both the galaxies and intergalactic absorbers at that time. While existing experiments will make precise statistical measurements over the next decade, we argue that improved 21-cm analysis techniques – allowing imaging of the neutral gas itself – as well as improved theoretical models, are crucial for testing our understanding of this important era.

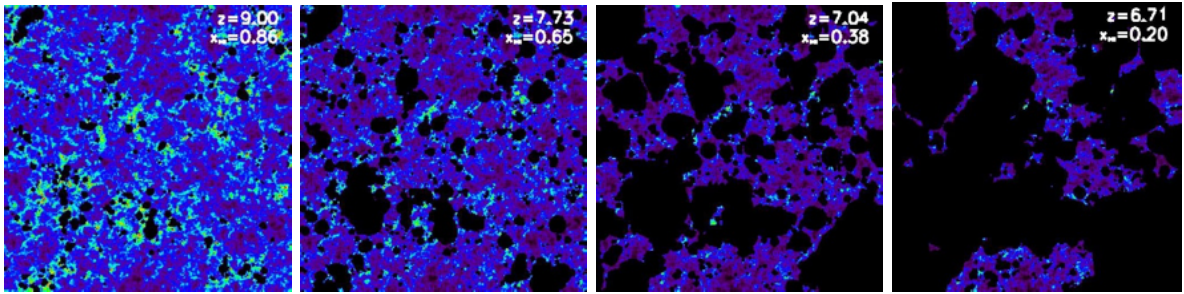

**Fig. 1**: **The 21-cm line provides a detailed history of reionization.** These slices through a semi-numeric simulation show the 21-cm signal expected at four different points during reionization (at ionized fractions $x_i$ = 0.86, 0.65, 0.38, and 0.20 from left to right). Black regions are ionized; the colorscale shows 21-cm intensity. This slice is 143 Mpc across. Courtesy A. Mesinger.

## I. INTRODUCTION

Reionization, when early generations of stars and (possibly) quasars ionized the intergalactic medium (IGM) about one billion years after the Big Bang, is one of the landmark events in the cosmic timeline. It is the most recent global phase transition in the Universe's history – the moment when galaxy formation affected every baryon in the Universe. It is accompanied by substantial photoheating that affects later generations of galaxies – including dwarf galaxies near the Milky Way. It reflects the early growth of the metagalactic radiation background, whose properties are important for the IGM to the present day. And, most importantly, it is the hallmark event of the first generations of galaxies, when those early sources transformed the Universe into the state we find it today. It is a sensitive probe of their properties and especially of their interactions with the IGM.

    Observations are beginning to probe the reionization era, a massive improvement from a decade ago – though there are still few quantitative constraints. Scattering of the cosmic microwave background by electrons after reionization provides an integrated constraint on the reionization history that indicates it ended relatively late ($z\sim6$) [1]. A decline in the apparent abundance of Ly-α emitting galaxies at $z\sim7$ may indicate that the Universe had a large neutral fraction at that point (e.g., [2-5]). Observations of Ly-α absorption toward high-redshift quasars also appear to indicate substantial reservoirs of neutral gas at $z\sim7$ (e.g., [6-7]).

    Though these observations paint a consistent picture, they do not yet provide a complete timeline of reionization. Because reionization depends on the integrated galaxy population, even its earliest stages are important probes of galaxy evolution. Moreover, the reionization *process* – its progress and morphology – is a sensitive probe of the interaction of those sources with the IGM. Ideally, we would like to watch the process unfold. The best way to do this is with the 21-cm line of neutral hydrogen, which could allow us to map the IGM throughout reionization, as illustrated in Fig. 1 (e.g., [8-10]).

    Such measurements have important implications for a wide range of astrophysical phenomena. The reionization history accessible via the 21-cm line will improve our interpretation of galaxy evolution (for example, in understanding the lack of recent star formation in many dwarf galaxies near the Milky Way as well as the properties of Lyman-α emitters at $z>5$), as a

baseline for understanding IGM physics at later times (in particular the thermal history and ionizing background), and cosmology (for which the ionization history is a "nuisance" parameter in CMB measurements; [11]).

In this white paper, we argue that measurements of the reionization history through the 21-cm line will open a new window onto the early Universe. Here we focus on measurements of the reionization process here; companion white papers by Furlanetto et al. and La Plante et al. place this signal in the context of other high-redshift probes, while Mirocha et al. describe how 21-cm measurements will shed light on the earliest luminous sources in the Universe.

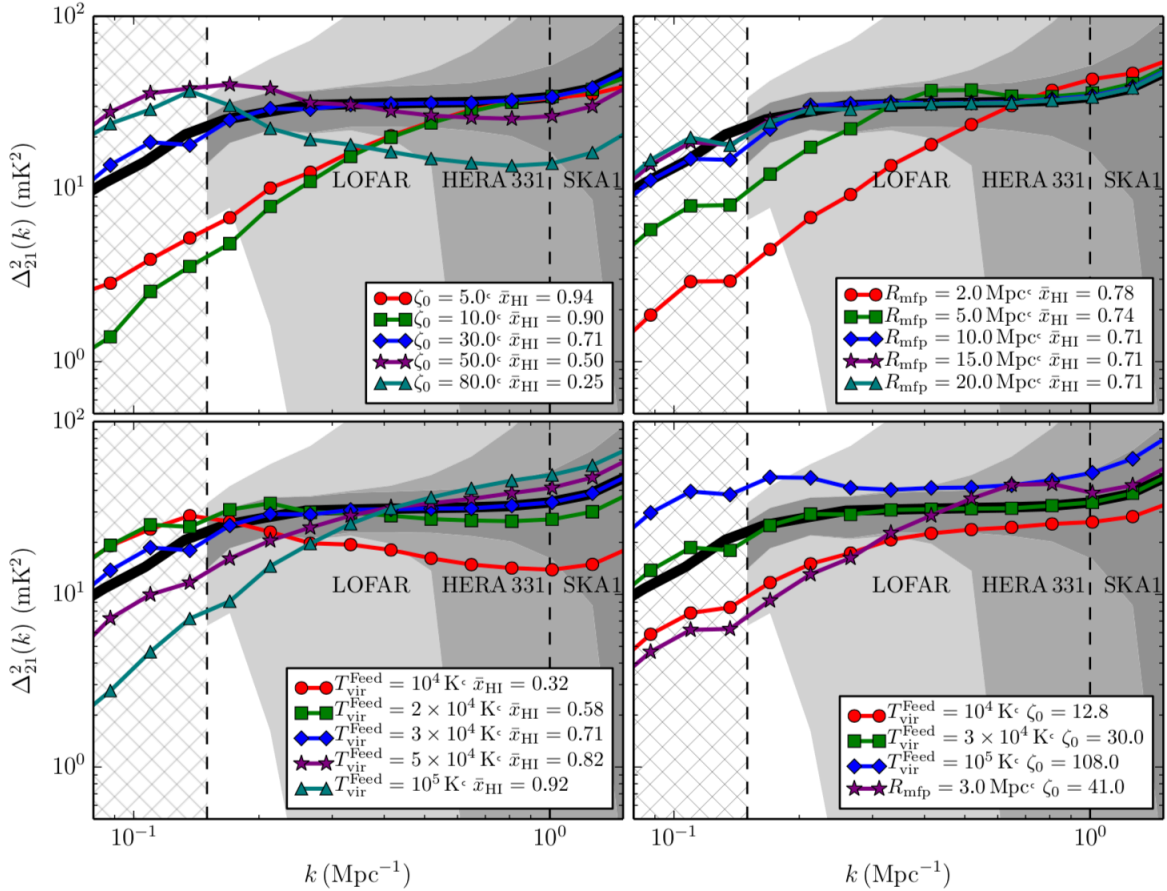

Fig. 2: **The 21-cm power spectrum is sensitive to the unknown properties of high-redshift sources.** In each panel, the curves show the power spectrum for different models of reionization. In all cases, the black curve is a fiducial model with $x_{HI}$=0.74 at z=9. Each panel varies different parameters: the ionizing efficiency (upper left), amount of IGM absorption (upper right), the minimum virial temperature of galaxies (lower left), and a variety of parameter choices that all produce $x_{HI}$=0.74. From [12].

## II. OPPORTUNITIES FOR THE NEXT DECADE

Low-frequency radio observatories are now finally reaching the stage where they can make these measurements. [13] reported the first detection of the sky-averaged 21-cm signal with the EDGES instrument. If confirmed, the detection requires substantial revision to our understanding of the Cosmic Dawn (e.g., [14-16]), but the implications of the measurement also point to the insights to be gained from 21-cm measurements. A number of other observatories are now operating (such as LOFAR [17]), under construction (such as the Hydrogen Epoch of Reionization Array, or HERA; [18]), or in the planning stages (such as SKA-Low).

The first generations of 21-cm observatories will attempt to measure the statistical properties of the 21-cm background – specifically, its power spectrum. Fig. 2 shows an example of the effectiveness of this probe [12]. It uses a simple "semi-numerical" model of the reionization process, with only three free astrophysical parameters, to demonstrate the sensitivity of the power spectrum to the underlying astrophysics. The upper left panel shows how the power spectrum changes with the overall ionizing efficiency of galaxies; the dashed lines demarcate the region accessible to near-future experiments, while the grey bands show projected errors. The power spectrum measures the overall ionization state of the Universe and can therefore trace the reionization history. This history is a crucial component of understanding the early Universe, and its measurement is the primary near-term goal of 21-cm telescopes like HERA.

The other panels in Fig. 2 show that the power spectrum contains even more information: the upper right panel shows how IGM absorption affects the power spectrum, while the lower left shows how the minimum size of galaxies affects it in this model. The lower right panel shows that, even when the ionized fraction is fixed, these astrophysical parameters cause measurable differences in the power spectrum.

More broadly, the topology of reionization – and therefore the 21-cm signal – is sensitive to many of the integrated properties of the source population, including the galaxy masses and ionizing efficiencies and the source spectra, which depends on whether quasars contribute to the process [39-40]. It is also sensitive to IGM absorption, which is likely dominated by small, dense pockets in the cosmic web that remain optically thick. These regions, known at later times as Lyman-limit systems, regulate the end of the reionization process, and understanding their evolution is critical to understanding how the ionizing background was established, which contains relics of reionization even after that process was complete [19-20].

Because of the rich astrophysics driving the 21-cm signal, interpreting these early statistical results will not be straightforward. Statistical measurements must be interpreted in comparison to theoretical models, which are simplified not only because of practical concerns (like computing power) but also because our knowledge of the source population and IGM is so limited at these times [12, 21]. While careful statistical analysis of the power spectrum can, for example, tightly constrain the parameters of a simple galaxy formation model as shown in Fig. 2, testing the assumptions of both galaxy formation and IGM models – or, better, making model-*independent* measurements – will in turn require more advanced observations. The recent EDGES observation serves as a good example of the challenge: the measurement did not match existing models of galaxy formation. Interpreting it therefore requires the introduction of new source populations (e.g., [15,22]): even the first 21-cm observation outstripped existing models, demonstrating how hard theory must work in order to keep up with the real world!

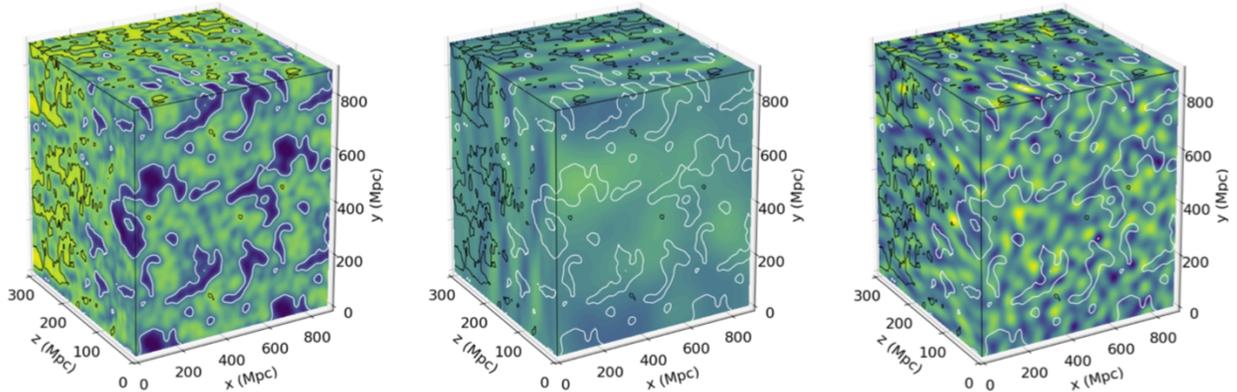

Fig. 3: **Improvements in 21-cm analysis pipelines are essential for testing reionization models.** The left panel shows a simulation of reionization, with the blue regions ionized and the colorscale showing 21-cm intensity. The white and black contours outline ionized and neutral regions, respectively – also shown in the other panels. The center panel shows a simulated HERA image of the same region, assuming existing analysis procedures. The right panel shows the same but assumes improved foreground removal. Note how the resolution and fidelity of the observation improves significantly because of the pipeline. Credit: A. Beardsley.

### III. KEY SCIENCE ADVANCES

Other statistics will undoubtedly prove useful in this regard [23-25], but the ultimate goal is to make high signal-to-noise images of the 21-cm background, which can probe the detailed physics in a model-independent manner.

The fundamental challenge in reaching this goal is to separate the (weak) cosmological signal from astrophysical foregrounds, which are four or more orders of magnitude brighter. "Foreground avoidance" techniques isolate this contamination within a data cube and allow statistical measurements and even some "imaging" (see the center panel of Fig. 3 for an example) [26-27]. However, foregrounds contaminate most of the interferometric data, so existing techniques provide only low signal-to-noise and poor fidelity, even on coarse angular scales of tens of arcminutes. The development of algorithms to remove more of the foregrounds from the contaminated region (known as the *wedge* in Fourier space), and hence retain more of the data, is essential for probing the details of reionization (see the right panel of Fig. 3).

**Imaging, or more advanced statistical measurements, will enable more precise tests of the reionization process and source properties.** Some of these tests are best made in comparison to the galaxy population and are described in another white paper by Furlanetto et al. But many can be made with the 21-cm data alone. For example, some simple tests enabled by imaging include: (1) Measuring the sizes and environments of the last remaining neutral regions will establish how reionization ends and show how the cosmic web develops; the fluctuations in the ionizing background induced by these regions persist after reionization is likely over [19-20, 28-29]. (2) The size distribution of ionized regions measures the details of the source-absorber relationship [12,30-31, 39-40] (3) Topological measurements of the ionized structure provide a

probe of source clustering [32-34]. (4) Identifying large ionized regions will offer unprecedented direct constraints on quasar lifetime and geometry [35-36]. (5) The identification of individual ionized regions will greatly enhance our trust in the 21-cm measurements.

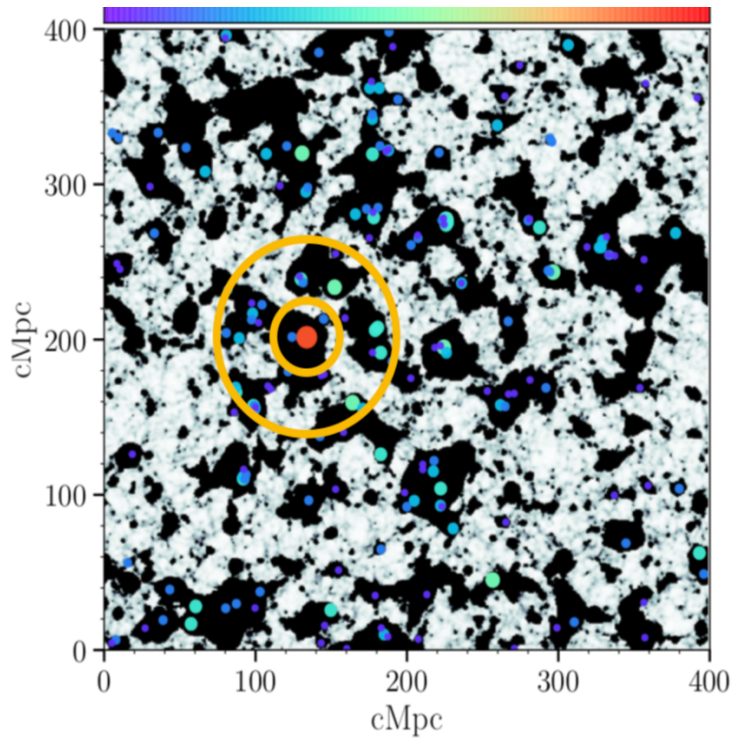

Fig. 4: **The identification of HII regions around luminous quasars during the epoch of reionization will provide unprecedented insight into their properties.** Here we a slice through a semi-numerical simulation near the midpoint of reionization (where black areas are ionized). Colored dots identify massive halos. If a luminous quasar turned on in the red halo, the extent of the ionized region would provide a direct measurement of its lifetime (the yellow circles show approximate radii after $10^{6.5}$ or $10^8$ years), while the region's shape (measured in three dimensions thanks to redshift information) would constrain the emission pattern. Adapted from [6].

Of course, these **precision observations will also challenge our theoretical models of both galaxy formation and reionization.** We therefore also require continued development of those models. These include numerical simulations of galaxy formation in this high-redshift regime as well as of the reionization process. Because the foreground problem is less difficult at high frequencies, imaging will likely first constrain the end of the reionization process. Recent observations [19] suggest that this epoch requires more sophisticated models of the cosmic web and ionizing background than currently available [20,28]. Given the many uncertainties in both galaxy formation and IGM absorption, rapid approximate techniques (including both analytic models [37] and "semi-numeric" simulations like 21cmFAST; [38]) will also be required to interpret the results and identify key observational strategies. A set of diverse, physically-motivated models, combined with detailed 21-cm observations and "orthogonal" constraints from other techniques, provides our best bet for understanding the early universe.

IV.     CONCLUSIONS

The 21-cm line has enormous potential to open the reionization era to detailed study. Statistical measurements are now on the horizon, but these will provide only the first steps in a detailed understanding of the reionization process. Continued development of analysis techniques for 21-cm measurements, opening access to the imaging regime, as well as improved theoretical models of the era, will enable huge strides in our understanding of the first galaxies.